\documentclass[english]{article}
\usepackage[T1]{fontenc}
\usepackage[latin9]{inputenc}
\usepackage{graphicx}
\usepackage{babel}
\usepackage{amsmath}
\usepackage{color}

\makeatletter

\newcommand{\bsubs}{\begin{subequations}}
\newcommand{\esubs}{\end{subequations}}

\newcommand{\be}{\begin{equation}}
\newcommand{\ee}{\end{equation}}

\newcommand{\bea}{\begin{eqnarray}}
\newcommand{\eea}{\end{eqnarray}}

\begin{document}

\title{Kenneth Geddes Wilson, 1936-2013, An Appreciation\\
 }
\author{ Leo P. Kadanoff\footnote{e-mail:  leop@UChicago.edu} \\
\\
The Perimeter Institute,\\
Waterloo, Ontario, Canada \\
\\
and \\
\\
The James Franck Institute\\
The University of Chicago
\\ Chicago, IL USA}

\maketitle
\begin{abstract}
Kenneth G. Wilson made deep and insightful contributions to statistical physics, particle physics, and related fields.  He was also a helpful and thoughtful human being.
\end{abstract}

\newpage{}
Ken's accomplishments in physics started with contributions to particle physics\cite{PP1,PP2,PP3,RG0}, work done mostly at Cal Tech, Harvard, SLAC, CERN, and Cornell.  In about 1970, he realized the close connection between approaches that could be used in particle physics and those that could be  applied  to the statistical mechanics of phase transitions.  The latter area had a rich experimental, anaytical and numerical tradition, so that many theoretical results could be verified or disproven by comparison with known results.  I shall mostly focus here upon the statistical work, since I am not fully qualified to evaluate contributions to particle theory.
 
The focus upon phase transitions produced a set of  remarkably important contribution to physics via the synthesis and construction  of the modern renormalization group technique\cite{RG0,RG1,RG2,RG3}.   Wilson used the threads of many previous authors, but wove them together with several essential pieces of his own invention.  The first of these new contributions was his inclusion of all the different  interaction terms consonant with the symmetries of the problem under consideration.   In addition he made innovative use of the fixed point concept to tie together the entire analysis.

However,  new tools form but a small part of his contributions to renormalization.  Prior to Wilson's accomplishments, renormalization, even as applied by such greats as Feynman, Tomonaga, and Schwinger, 
was a method of dubious validity and only partial respectability.  Wilson first showed how to put it on a much firmer mathematical footing. Then,  his $\epsilon$-expansion work with Michael Fisher\cite{WF}   demonstrated the correctness of the renormalization idea by using it to determine, with remarkable accuracy, the primary experimental facts about the liquid-gas critical point.   That work included a technical innovation of substantial importance, the idea of treating problems by considering their continuous variation with spatial dimension, and particularly of looking for dimensions where the problem might be simplified\cite{epsilon}.

After its modern construction by Wilson and others, the renormalization group has  appeared in thousands of papers devoted to the development of the understanding of physical, social, biological and financial systems.  However, renormalization is substantially more than a technical tool.  It is primarily a method for connecting the behavior at one scale to the phenomena at a very different scale.  It serves for example, to connect the physics at the scale of an atom with the observed macroscopic properties of materials.   One might argue, and I believe that argument, that the connection among "laws of nature" at different scales of energy, length, or aggregation is the root subject of physics\cite{scales}.  One would then argue that Wilson has provided us with the single most relevant tool for understanding physics.  

Ever since the early 1970s, the tools and concepts put forward by Wilson have formed the very basis of particle physics, field theory, and condensed matter physics. These concepts include  fixed points, couplings that vary with scale, variation of physical properties with spatial dimension, description of couplings via anomalous dimension, qualitative variation in properties as a consequence of phase transitions,  and topological descriptions of excitations. 

That is, of course, Ken's main accomplishment.  But it is not all.  He has helped add to our profession a new style of work and of thinking.  When he said that we should do renormalization by looking at all the coupling that might arise with a given symmetry, he was asking for something impossible.  Nobody can keep track of hundreds of different couplings.  But a computer can do so....     Wilson was implicitly suggesting we could gain understanding by asking what a computer can do, and then what we ourselves might do by either using or emulating computers.  This mode of thinking was present  in the first stat. mech. renormalization papers\cite{RG1,RG2} but even more so in the work that Ken did to understand the Kondo problem\cite{K}, a situation in which a single nuclear spin interacts with all the mobile electronic spins in a metal.  This situation was analyzed using a numerical version of renormalization,  involving hundreds of couplings.  This style of calculation was carried forward by Steve White\cite{White} and others to produce a numerical renormalization technique, informed by ideas of information theory\cite{Vidal}.  This kind of work has become a productive subfield of physics.

Ken's engagement in the use of computers naturally developed into an interest in development of methods for supercomputer calculations, which then included a sponsorship of proposals for supercomputers and supercomputer centers, and then morphed into the design of a complex of computer programs which might make possible flexible use of very large computers.  The computer-work was, in part, carried out with Ken's wife Alison Brown, who is a computer scientist.

Another subfield engendered by Wilson's work\cite{lattice,lattice1}  is lattice gauge theory.  In this area, one constructs computer simulations of a field theory, often quantum chromodynamics, by doing renormalization calculations in which the field variables are placed upon a lattice.  Wilson did the first calculation of this kind\cite{lattice}, which has now been followed by many others.    The result were concrete realizations of quantum chromodynamics, which have enabled the calculation of many of the parameters of this theory.

In addition,Wilson has created several  important analytical innovations.   The short distance expansion\cite{ope} or operator product expansion\cite{ope1} provides an opening to an algebra formed by the fluctuating operators in field theory or statistical mechanics.  This approach then gives us the possibility of analytic understanding of field theories. One important result is the descriptions of  a wide variety of two-dimensional critical behaviors\cite{cft}.    

Another invention, the Wilson loop, was among the first of many topological innovations that have enriched particle physics and field theory.    Understanding this loop permitted important insights into why quarks were crucial for hadron behavior, but at the same time very hard to observe\cite{lattice}.

Even given this wonderful breadth of accomplishment within physics, the renormalization effort stands out as Ken's most important accomplishment.  Before Ken's work, P.A.M. Dirac could say of renormalization theory "I might have thought that the new ideas were correct if they had not been so ugly."   Afterwards renormalization became simply a tool for connecting different theories.

Ken was unfailingly generous and broadly helpful to the rest of us working in his area.  One evidence of generosity is the very careful credit given to previous work and workers  in Ken's papers. (See, for example, the extensive references in papers\cite{RG0,RG1,RG3}.)   Moreover, I remember with some gratitude the result of a trip to Cornell which I took to help me understand how fermions might be included in a paper I was writing about quarks and strings.  My visit elicited from Ken a tutorial on Grassman variables, which I then used in my paper\cite{LPKxxx}.  

The brilliance of Kenneth G. Wilson was dazzling, but he never tried to outshine those about him.   He was all quiet competence\cite{obit}.

His PhD students include Roman Jackiw, Michael Peskin,  Paul Ginsparg,  Steven R. White, Serge Rudaz and Steve Shenker, who worked under both Ken and Ken's coworker John Kogut.

\section*{Acknowledgements} I have had helpful suggestions from Michael Fisher, Ruth Kadanoff,  Mitchell Feigenbaum, Gloria Lubkin, Steve White, Silvan Schweber, and Stephen Shenker.  This work was was supported in part by the University of Chicago MRSEC program under NSF DMR-MRSEC  grant number 0820054.  It was also supported in part by the Perimeter Institute,  which is supported by the Government of Canada through Industry Canada and by the Province of Ontario through the Ministry of Research and Innovation.

  \end{document}